\documentclass[]{aa}
\usepackage{natbib}
\usepackage{graphicx}
\usepackage{ulem}
\usepackage{hyperref}
\usepackage{color}
\usepackage{soul}

\begin{document}

\title{A distinct magnetic property of the inner penumbral boundary \\
III. Analysis of simulated sunspots}
\titlerunning{A distinct magnetic property of the inner penumbral boundary. III.}
\author{Jan Jur\v{c}\'{a}k
        \inst{1}
        \and
        Markus Schmassmann
        \inst{2}
        \and
        Matthias Rempel
        \inst{3}
        \and
        Nazaret Bello Gonz\'alez
        \inst{2}
        \and
      	Rolf Schlichenmaier
        \inst{2}
       }

\institute{Astronomical Institute of the Czech Academy of Sciences, Fri\v{c}ova  298, 25165 Ond\v{r}ejov, Czech Republic
\and 
  Leibniz-Institut f\"{u}r Sonnenphysik (KIS), Sch\"{o}neckstr. 6, 79104 Freiburg, Germany
  \and
  High Altitude Observatory, NCAR, P.O. Box 3000, Boulder, CO 80307, USA
  }

\date{Received 11 December, 2014; accepted }

\abstract
{The analyses of  sunspot observations revealed a fundamental magnetic property of the umbral boundary, the invariance of the vertical component of the magnetic field.}
  {We aim to analyse the magnetic properties of the umbra-penumbra boundary in simulated sunspots and thus assess their similarity to observed sunspots. Also, we aim to investigate the role of plasma $\beta$ and the ratio of kinetic to magnetic energy in simulated sunspots on the convective motions as these quantities cannot be reliably determined from observations.}
 {We use a set of non-grey simulation runs of sunspots with the MURaM code. The setups differ in terms of subsurface magnetic field structure as well as the magnetic field boundary imposed at the top of the simulation domain. These data are used to synthesise the Stokes profiles that are then degraded to the Hinode spectropolarimeter-like observations. Then, the data are treated like real Hinode observations of a sunspot and magnetic properties at the umbral boundaries are determined.}
{Simulations with potential field extrapolation produce a realistic magnetic field configuration on their umbral boundaries. Two simulations with potential field upper boundary, but different subsurface magnetic field structures, differ significantly in the extent of their penumbrae. Increasing the penumbra width by forcing more horizontal magnetic fields at the upper boundary results in magnetic properties that are not consistent with observations. This implies that the size of the penumbra is given by the subsurface structure of the magnetic field.  Namely, the depth and inclination of the magnetopause shaped by the sunspot flux rope expansion with height. None of the sunspot simulations is consistent with observed properties of the magnetic field and direction of the Evershed flow  at the same time. Strong outward directed Evershed flows are only found in setups with artificially enhanced horizontal component of the magnetic field at the top boundary that are not consistent with the observed magnetic field properties at the UP boundary. We want to stress out that the  `photospheric' boundary of simulated sunspots is defined by a magnetic field strength of equipartition field value.}
{}
  

\maketitle

%
%
\section{Introduction}

Recent analyses of spectropolarimetric observations proved the importance of the vertical component of the magnetic field ($B_\mathrm{ver}$) on the stability of sunspot umbrae. In the initial paper of \citet{Jurcak:2011}, the author showed that $B_\mathrm{ver}$ is the only magnetic field property that is constant on the umbral boundaries of a small sample of symmetric sunspots that were analysed. The results did not provide any conclusive insight whether or not the constant $B_\mathrm{ver}$ values found for each umbra depend on its size.

Using a unique dataset capturing the formation of a penumbral segment at the proto-spot boundary, \citet{Jurcak:2015} found that the forming umbra/penumbra boundary is migrating towards the umbral core into region with stronger and more vertical magnetic field until it reaches its stable position at a location where $B_\mathrm{ver}$ is comparable to the values found in stable sunspots in \citet{Jurcak:2011}. The authors concluded that in regions with $B_\mathrm{ver} < B_\mathrm{ver}^\mathrm{stable}$ the penumbral mode of magneto-convection takes over the umbral mode. This was further confirmed by an analysis of a pore with $B_\mathrm{ver} < B_\mathrm{ver}^\mathrm{stable}$ that is completely colonised by a growing penumbra \citep{Jurcak:2017}. 

These results motivated a Bayesian statistical analysis of 114 umbral boundaries that showed no dependence of $B_\mathrm{ver}$ averaged along the umbral boundary based on continuum intensity ($I_c = 0.5 I_c^\mathrm{QS}$) on spot size \citep{Jurcak:2018}. In case of the Hinode spectropolarimetric (SP) observations \citep{Kosugi:2007, Tsuneta:2008} and the employed inversion method, the most likely value of $B_\mathrm{ver}$ is 1867~G and the value is with 99\% probability in the range from 1849~G to 1885~G. 
\citet{Schmassmann:2018} analysed data from the HMI instrument \citep{Schou:2012} on-board the SDO satellite to investigate the evolution of $B_\mathrm{ver}$ during a stable phase of one long-lived sunspot. The results confirmed that $B_\mathrm{ver}$ is the only constant parameter on the umbral boundary as both magnetic field strength, $|B|$, and inclination, $\gamma$, vary with the lifetime of the sunspot. The authors found a stable $B_\mathrm{ver}$ value of 1693~G, where the difference compared to the $B_\mathrm{ver}$ value obtained with Hinode SP data is plausibly explained by the different spectral and spatial resolution. Furthermore, they confirmed quantitatively that contours based on $B_\mathrm{ver}$ match the umbral boundary defined by continuum intensity better than those based on $|B|$ and $\gamma_\mathsc{lrf}$.

Another observational analysis of the $B_\mathrm{ver}$ evolution on the umbral boundary of a decaying sunspot was done by \citet{Benko:2018}. The results show that in the decaying phase, the continuum intensity boundary of the umbra ($0.5 I_\mathrm{c}^\mathrm{QS}$) does not match the boundary based on $B_\mathrm{ver}$ and this discrepancy is observed already at a phase when the sunspot look morphologically regular. The values of $B_\mathrm{ver}$ are smaller than $B_\mathrm{ver}^\mathrm{stable}$ in the umbra and it is thus unstable and prone to be colonised by more vigorous modes of magneto-convection. 

The observational evidence for the importance of $B_\mathrm{ver}$ for a stable umbra is well supported by the above-mentioned analyses. However, the interpretation of SP observations does not allow us to investigate quantities like plasma $\beta$ and the ratio of kinetic to magnetic energy there, as inversions do not provide us with reliable densities (inversion codes assume a hydrostatic equilibrium) and we cannot compare physical parameters at the same geometrical height (inversion codes operate in optical depth scale for each pixel separately). This motivated us to analyse simulations of sunspots. First, we investigate whether or not the relation between intensity and $B_\mathrm{ver}$ boundary is present in the simulations and compare the properties of observed and simulated sunspots. Second, using the physical parameters from the simulated data cubes, we investigate the behaviour of plasma $\beta$ and the ratio of kinetic to magnetic energy within the simulated sunspots attempting to enhance our knowledge of penumbral and umbral magnetoconvection.

From a theoretical perspective, \citet{Chandrasekhar:1961} pointed out that the vertical component of the magnetic field is the key parameter for the stability against overturning convection in the presence of magnetic field, when the Boussinesq approximation is valid. The horizontal component of the magnetic field just determines the shape of the convective cell. \citet{Mullan:2019} related the observed constant $B_\mathrm{ver}$ value to the analysis of \citet{Gough:1966}, who extended Schwarzschild's convective stability criterion for compressible gases to include the stabilising effect of the magnetic field. This Gough-Tayler criterion shows where convective motions are inhibited and is expressed in the simplest form as
\begin{equation}
    \frac{1}{\Gamma_1} - \frac{\mathrm{d} \ln \rho}{\mathrm{d} \ln p} < \frac{B^2_\mathrm{ver}}{B^2_\mathrm{ver} + 4 \pi \Gamma_1 p},
    \;\textrm{with}\;\Gamma_1=\left(\frac{\mathrm{d} \ln p}{\mathrm{d} \ln \rho}\right)_\mathrm{ ad}, 
\end{equation}
where $\Gamma_1$ is Schwarzschild's first adiabatic exponent, $\rho$ is the density and $p$ is the pressure. The relevance of this theoretically derived stability criterion in the analysed MHD simulations of sunspots is currently under investigation. 

\section{Simulations and data processing}
\label{observations}

We use two different types of MHD simulations of sunspots carried out with the radiative MHD code MURaM \citep{Vogler:2005, Rempel:2009a}. All the analysed simulations employed a non-grey radiative transfer and had horizontal and vertical grid size of 32~km and 16~km, respectively (in several hours of solar time before the time step we analyse). 

The first simulation (type I) is inspired by the approach of {\AA}ke Nordlund\footnote{On sunspots and penumbrae. Nordita Seminar on Sunspot formation: theory, simulations and observations, 2015, {\href{http://video.albanova.se/ALBANOVA20150313/video1.mp4}{Link}}}, where the flux tube is strongly compressed at the bottom boundary of the simulation domain. The initial magnetic flux throughout the whole box is $\Phi=10^{22}\,\textrm{Mx}$. We initialise this simulation with velocities, densities and internal energies of a quiet Sun simulation upon which a magnetic field is imposed. The vertical field is defined at the bottom boundary as
\begin{equation}
    B_\mathrm{ver,0}=B_0\exp\left(-\frac{x^2+y^2}{\pi \Phi/B_0}\right), 
\end{equation}
with $B_0=160\,\textrm{kG}$ and the following potential field condition gives the rest of the initial field
\begin{equation}
    \vec{B}=-\nabla\cdot\mathcal{F}^{-1}\left(\mathcal{F}(B_\mathrm{ver,0})
    \frac{\exp\left(-(z-z_0)|k|\right)}{|k|}\right),
\end{equation}
whereby $z_0=0$ is at the bottom of the box, $\mathcal{F}$ is the Fourier transform in $x$ and $y$ direction, and $|k|=\sqrt{k_x^2+k_y^2}$. The upper boundary condition guarantees that field above the upper boundary remains such a potential field \citep{Cheung:2006,Rempel:2012}. As the bottom boundary allows for vertical mass exchange, the field strength there relaxes from $160$~kG to about $30$~kG after the initial 6 hours of the simulation.

The second type of simulation (type II) contains four different runs that are described in detail in \citet{Rempel:2012}. The initial magnetic flux for all runs is $1.2 \times 10^{22}$~Mx. Type~II simulations have significantly lower magnetic field strength at the bottom boundary than type~I. At the time of the analysed type~II snapshots, the field strength is around $7$~kG at the bottom boundary. The simulation setup is such that the upper boundary condition can force the magnetic field to be more inclined to enhance the sunspot penumbra and the magnetic field can thus be non-potential. The deviation from the potential case is controlled by the $\alpha$ parameter that is equal to 1 (potential case), 1.5, 2, and 2.5. Fig.~2 in \citet{Rempel:2012} shows how the width of the penumbra increases with increasing $\alpha$.

\begin{figure}[!t]
 \includegraphics[width=\linewidth]{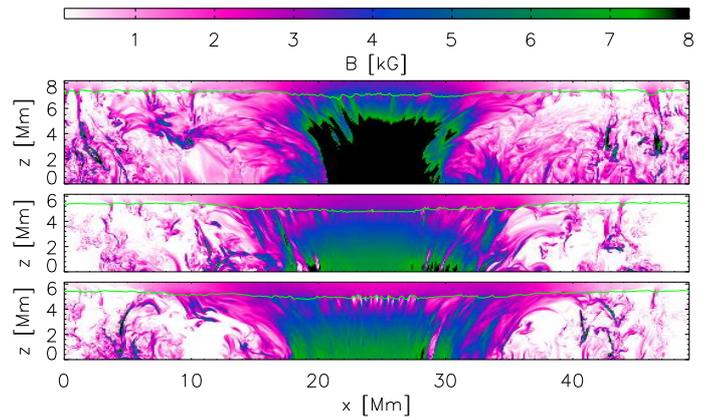}
 \caption{Vertical cuts through the flux tube of the type I (top) and type II ($\alpha = 1$, middle; $\alpha = 2.5$, bottom) simulations showing the magnetic field strength. The green lines mark the optical depth unity ($\tau = 1$).}
 \label{mag_cut}
\end{figure}

In Fig.~\ref{mag_cut}, we show the magnetic field strength along the vertical cut through the entire simulation domain for type I and II ($\alpha = 1$ and $2.5$) runs. This comparison illustrates the consequences of the above mentioned setups of the different simulations, i.e., the opening of the magnetic flux rope with height and the variation of field strength with height.

As the output of the MHD simulations, we have data cubes of temperature, density, gas pressure, and vector of velocity and magnetic field. Using a routine "optical.x" from the SIR code \citep[Stokes Inversion based on Response function,][]{Cobo:1992}, we transform pixel by pixel the geometrical height scale to optical depth scale and as a by-product we also obtain the electron pressure. Thus, we have all the necessary parameters to synthesise in every pixel the Stokes profiles for the \ion{Fe}{i} spectral lines around 630~nm that are observed by Hinode SP. For this step, we assume that there is no microturbulent velocity, the filling factor is unity, and there is no stray light. In the synthesis, we take into account the spectral point-spread function (PSF) of the Hinode SP. Note that we have to refine the vertical grid size of the simulation datacubes by using linear interpolation to correctly synthesise the Stokes profiles.

The resulting maps of intensities at all wavelengths and all Stokes profiles were then convolved with the Hinode spatial PSF \citep{Noort:2012}. The resulting data were then re-binned to the spatial sampling of the Hinode SP to mimic the observations where we assumed both the normal and fast scanning modes with pixel sizes of $0\farcs1486 \times 0\farcs16$ and  $0\farcs297 \times 0\farcs32$, respectively. As a last step in data degradation, we added the noise level estimated from actual Hinode SP observations ($2.4 \times 10^{-3}$ and $1.8 \times 10^{-3}$ of the quiet Sun continuum intensity, $I_\mathrm{c}^\mathrm{QS}$, for the normal- and fast-like scans, respectively). 

\begin{figure*}[!t]
 \sidecaption
 \includegraphics[width=0.7\linewidth]{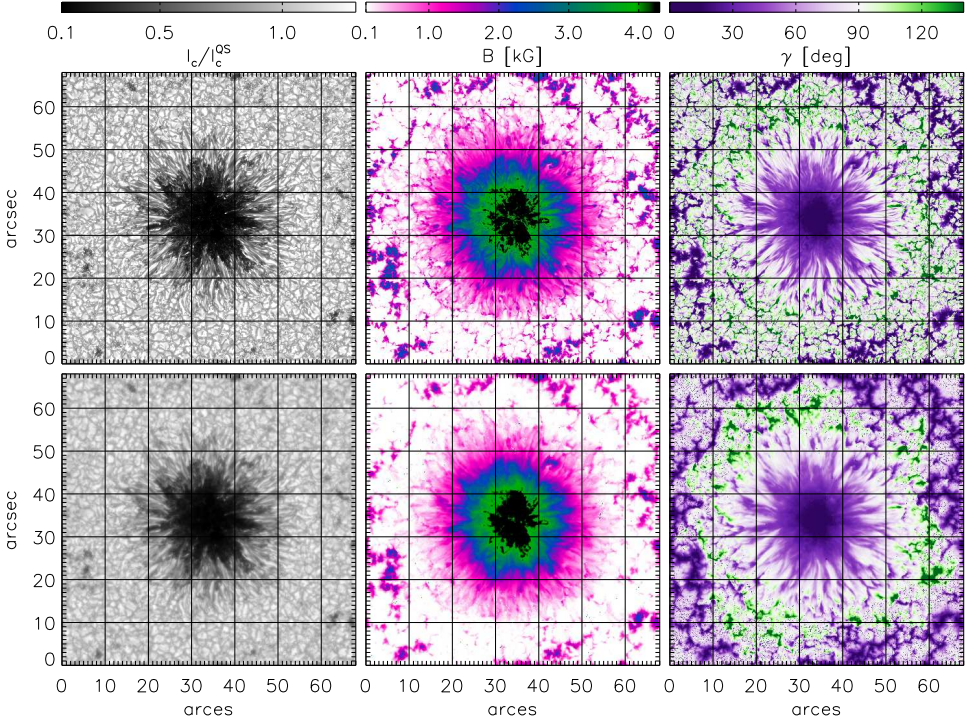}
 \caption{Maps from left to right: continuum intensity at 630.1~nm, magnetic field strength and inclination. Upper row corresponds to the resolution of the MHD simulation and values of physical parameters are taken directly from the simulation box (averaged values between $\log \tau = -1$ and $-2$). Bottom row corresponds to the results of the SIR inversion, i.e., the continuum intensity is taken from the fitted profile and magnetic field strength and inclination are high-independent values from the resulting model atmosphere.}
 \label{overview_maps}
\end{figure*}

\begin{figure*}[!t]
 \sidecaption 
 \includegraphics[width=0.7\linewidth]{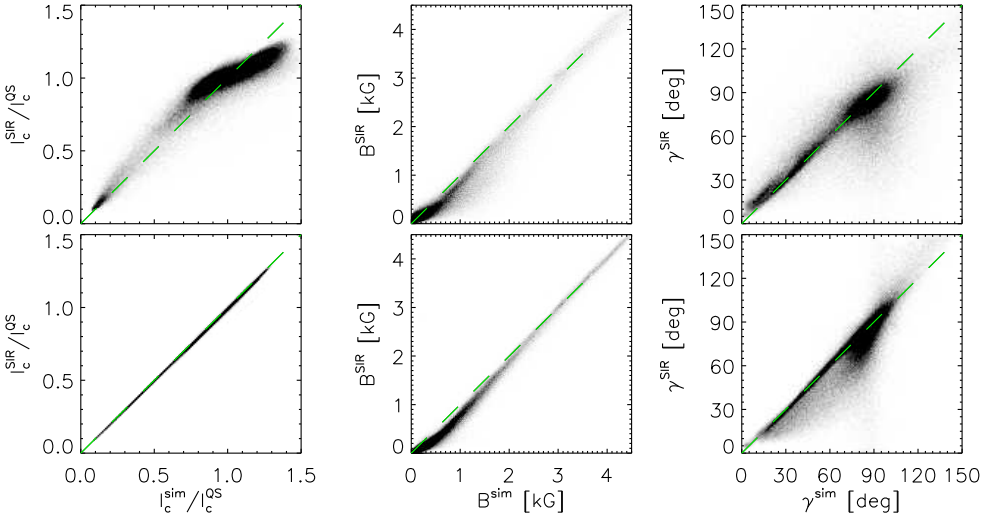}
 \caption{Scatter plots showing the accuracy of the inverted parameters with respect to the actual values in the simulation box (average values between $\log \tau = -1$ and $-2$). The values on the horizontal axis are taken directly from the simulation box in the upper row; in the bottom row these values are smoothed by the spatial PSF. The dashed green lines mark the one-to-one correspondence.}
 \label{scatter_plots}
\end{figure*}

Afterwards, we treated the degraded synthetic profiles like real Hinode SP observations. To determine the magnetic field properties, we used the inversion code SIR. The inversion setup is identical to the one we used in \citet{Jurcak:2018}:
\begin{itemize}
    \item temperature is allowed to change with height
    \item all other atmospheric parameters ($\vec{B},v_\mathrm{LOS}$) are constant
    \item the spectral PSF of Hinode SP is taken into account
    \item one-component model atmosphere with filling factor unity and no stray light 
    \item macro-turbulence is set to zero, micro-turbulence is a free parameter of the inversion
\end{itemize}

\begin{figure*}[!t]
 \sidecaption
 \includegraphics[width=0.77\linewidth]{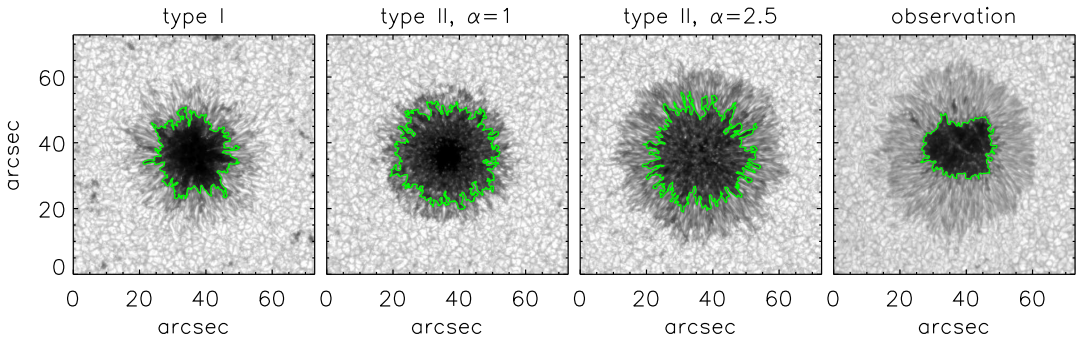}
 \caption{Continuum intensity maps at 630.1~nm for analysed snapshots of simulated sunspots (degraded to the Hinode-like SP observations) and for an observed sunspots with comparable $\Phi$. The green contours mark the UP boundary defined at $0.5 I^\mathrm{QS}_\mathrm{c}$.}
 \label{spots_int}
\end{figure*}

We assess the reliability of the inversion by comparing the original values from the simulation to the results obtained by the SIR code. In Fig.~\ref{overview_maps}, we show the original full-resolution maps from the simulation box (upper row) and the results of the inversion of the synthetic profiles degraded to the resolution of the Hinode SP normal-like scan (bottom row). Instead of temperature maps, we show the continuum intensity maps. The visual difference is caused mostly by the spatial smoothing induced by the Hinode spatial PSF. This is apparent from Fig.~\ref{scatter_plots}, where we compare the actual values from the simulation box (horizontal axes) with those resulting from the inversion (vertical axes). When these values are compared directly (upper row), the scatter plots show more discrepancies from one-to-one correspondence than when we use for comparison values from the simulation box smoothed by the spatial PSF (bottom row). These plots imply that our simplified inversion scheme retrieves the physical parameters of the simulation reliably and we can use them for comparison with physical parameters retrieved from observations of actual sunspots. Assuming that the MHD simulation produces a solar atmosphere with a complexity comparable to the actual solar atmosphere, the scatter plots shown in the upper row of Fig.~\ref{scatter_plots} illustrate the actual errors of plasma parameters in the observational studies of umbral boundaries that are based on Hinode SP data \citep{Jurcak:2011, Jurcak:2017, Jurcak:2018}.

\begin{figure*}[!t]
 \sidecaption
 \includegraphics[width=0.77\linewidth]{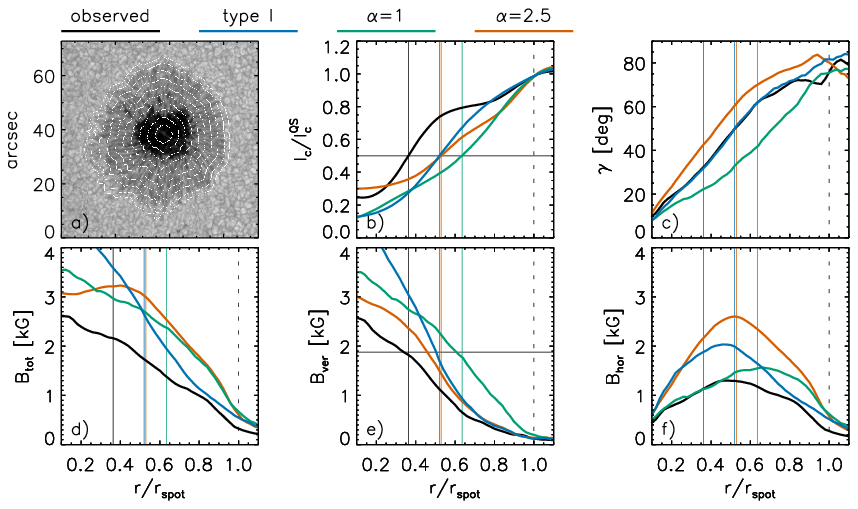}
 \caption{Radial profiles of continuum intensity ($I_\mathrm{c}$, b), magnetic field inclination ($\gamma$, c), total magnetic field strength ($B_\mathrm{tot}$, d), vertical magnetic field strength ($B_\mathrm{ver}$, e), and horizontal magnetic field strength ($B_\mathrm{hor}$, f). The continuum intensity map of the observed sunspot with contours marking relative radial positions from 0.1 to 1 is shown in panel (a). The solid vertical lines mark the UP boundaries, the dashed vertical lines mark the penumbra -- quiet Sun boundary. The horizontal lines in (b) and (e) mark the $0.5 I^\mathrm{QS}_\mathrm{c}$ and 1876~G, respectively.}
 \label{radial_profile}
\end{figure*}

\section{Comparison of synthetic and real sunspots}
\label{results}

For each of the analysed simulations, we computed the total magnetic flux of the sunspot area using the magnetic field strength and inclination resulting from the inversion. Type~I simulation has a magnetic flux ($\Phi$) of $7.7 \times 10^{21}$~Mx and all type~II simulations have $\Phi \sim 1.1 \times 10^{22}$~Mx. From the database of sunspots we analysed in \citet{Jurcak:2018}, we choose a symmetric sunspot with comparable $\Phi$ of $8.5 \times 10^{21}$~Mx for comparison. In Fig.~\ref{spots_int}, we compare the intensity maps of simulated sunspots with an actual Hinode SP observation. Visually,  the type II run with $\alpha = 2.5$ resembles the most the observed sunspot in terms of penumbral width, brightness and morphology, although this simulation shows too much fine structure in the umbra.

Similarly to our observational analyses, we define the umbra-penumbra (UP) boundary at 50\% of the quiet Sun continuum intensity ($0.5 I^\mathrm{QS}_\mathrm{c}$). These boundaries are marked in Fig.~\ref{spots_int} by green contours. Along these lines, we computed the mean magnetic field properties of the analysed snapshots and compare them to the observed sunspot in Table~\ref{UP_properties}. 

\begin{table}
\caption{Magnetic properties on the UP boundaries of simulated and observed sunspots}
\label{UP_properties}
\centering 
\begin{tabular}{l  c c c }
sunspot & $B$ [kG] & $\gamma$ [deg] & $B_\mathrm{ver}$ [kG] \\
\hline 
type I & $2.49 \pm 0.33$ & $49 \pm 5$ & $1.64 \pm 0.33$ \\
type II $\alpha=1$ & $2.33 \pm 0.26$ & $44 \pm 5$ & $1.68 \pm 0.31$ \\
type II $\alpha=1.5$ & $2.64 \pm 0.25$ & $58 \pm 6$ & $1.39 \pm 0.32$ \\
type II $\alpha=2$ & $2.73 \pm 0.26$ & $60 \pm 5$ & $1.36 \pm 0.33$ \\
type II $\alpha=2.5$ & $2.85 \pm 0.26$ & $62 \pm 6$ & $1.33 \pm 0.33$ \\
observed & $2.23 \pm 0.15$ & $33 \pm 3$ & $1.85 \pm 0.12$ \\
\hline
\end{tabular}
 \end{table}
 
The range of values of $B$ and $\gamma$ on the UP boundaries of observed sunspots can be found in Fig.~2 in \citet{Jurcak:2018}. Magnetic field strength can reach values from roughly 1.8~kG to 2.6~kG where the larger spots have stronger field on their UP boundaries. Magnetic field inclination can reach values from roughly $15^\circ$ to $45^\circ$ where the more inclined field is found in larger sunspots. Sunspots with $\Phi \sim 10^{22}$ (comparable to those of simulated sunspots) have $B \sim 2.2$~kG and $\gamma \sim 35^\circ$, i.e., the field is weaker and more vertical than those found on UP boundaries of simulated spots. 

Only simulations with potential field extrapolations (type I and type II with $\alpha = 1$) have values of $B$ and $\gamma$ on their UP boundaries that can be found in observed sunspots, but these have higher magnetic flux. Type II simulations with higher $\alpha$ have unrealistic horizontal magnetic fields on their UP boundaries that were never observed in sunspots and also the magnetic field strength is higher than in the case of observed sunspots. 

We know from observations that the $B_\mathrm{ver}$ on the UP boundaries of observed sunspots is independent on their size and thus on their magnetic flux. All the simulated sunspots have weaker $B_\mathrm{ver}$ than the observationally found $B_\mathrm{ver}^\mathrm{stable}$ of 1.87~kG. The most realistic are the simulations with potential field extrapolation that have $B_\mathrm{ver} \sim 0.2$~kG lower than $B_\mathrm{ver}^\mathrm{stable}$. 

\begin{table}
  \caption{Structural properties of the UP boundary}
\label{fractal_dim}
\centering 
\begin{tabular}{l  c c }
sunspot & fractal dimension & position ($r/r_\mathrm{spot}$) \\
\hline
type I & 1.20 & 0.52\\
type II $\alpha=1$ & 1.28 & 0.64 \\
type II $\alpha=1.5$ & 1.28 & 0.63 \\
type II $\alpha=2$ & 1.30 & 0.56 \\
type II $\alpha=2.5$ & 1.32 & 0.53 \\
observed & 1.08 & 0.36 \\
smooth circle & 1.01 & -- \\
\hline
\end{tabular}
 \end{table}

\begin{figure*}[!t]
 \sidecaption
 \includegraphics[width=0.7\linewidth]{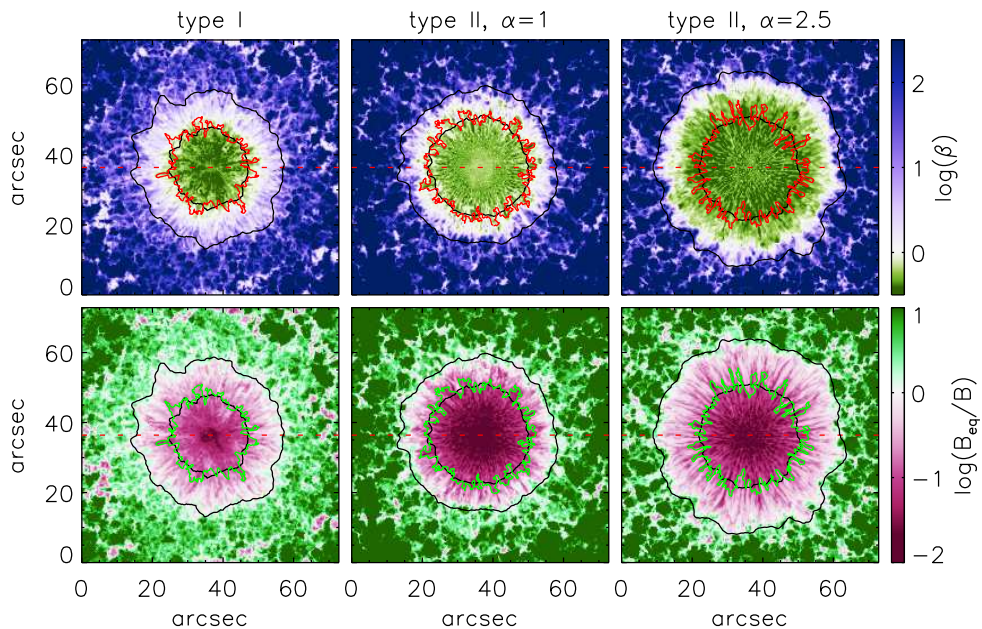}
 \caption{Maps of $\log(\beta)$ (top row) and $\log(B_\mathrm{eq}/B)$ (bottom row) at the continuum formation height ($\tau_{500}=1$) for different simulation runs. The red (top) and green (bottom row) contours mark the UP boundaries defined at $0.5 I^\mathrm{QS}_\mathrm{c}$. The black contours mark the position of the outer sunspot boundaries and the mean position of the UP boundaries that are defined for the purposes of azimuthal averaging (see Fig.~\ref{radial_profile}). The red dashed lines through the middle of the spots mark the position of the cuts shown in Fig.~\ref{forces_cuts}.}
 \label{forces_tau1}
\end{figure*}

For all physical parameters on the boundaries of simulated sunspots, we find significantly larger standard deviations than in the case of observed sunspot (the $\pm$ values in Table~\ref{UP_properties}). This is caused by the shape of the boundaries that are more corrugated in the case of the simulations (see the complexity of the green contours in Fig.~\ref{spots_int}). To quantify this property, we computed the fractal (Minkowski--Bouligand) dimensions of the UP boundary. The results are shown in Table~\ref{fractal_dim}, where we included a smooth circular boundary for comparison.
 
In Fig.~\ref{radial_profile}, we compare the radial profiles of various parameters of the simulated and observed sunspots. To do so, we defined the outer penumbral boundary as $0.99 I^\mathrm{QS}_\mathrm{c}$, where we smoothed the $I_\mathrm{c}$ with a box function $5" \times 5"$ and the sunspot centre as the centre of gravity of the umbra. Then, we computed for each pixel in the field of view its relative radial distance between the outer penumbral boundary and the sunspot centre. Then, we averaged pixels with the same relative radial distance over the position angle. Examples of iso-contours of relative radial distance are shown in Fig.~\ref{radial_profile}a.

The radial profiles of intensity allow us to determine the mean position of the UP boundary, i.e., $r/r_\mathrm{spot}$ where $I_\mathrm{c}/I^\mathrm{QS}_\mathrm{c} = 0.5$. In the case of the observed sunspot used for comparison in this paper, the UP boundary is located at $r/r_\mathrm{spot} = 0.36$. Also other observational studies show remarkable uniformity of UP boundary positions close to $r/r_\mathrm{spot} = 0.4$ \citep[e.g.,][]{Keppens:1996,cwp:2001, Mathew:2003, Borrero:2004, Bellot:2004, Sanchez:2005, Beck:2008, Borrero:2011}. In the case of simulated sunspots, the $r/r_\mathrm{spot}$ of the UP boundary is in all studied cases above 0.5, see Table~\ref{fractal_dim}.

Of particular interest to us is the radial profile of $B_\mathrm{ver}$ and its value on the mean position of the UP boundary. In the case of the observed sunspot, the resulting value is 1.82~kG, i.e., lower than the $B_\mathrm{ver}$ value found on the UP boundary as shown in Table~\ref{UP_properties}. On the other hand, the $B_\mathrm{ver}$ value on the mean position of the UP boundaries in simulated sunspots increased considerably compared to the actual UP boundary values. For the runs with potential field extrapolations, we obtain values comparable to the observed sunspot, i.e., 1.71~kG and 1.78~kG for type~I and type~II $\alpha = 1$, respectively. It is still too weak, but it confirms that the discrepancy between the observed and simulated sunspots is at least partially caused by the corrugated UP boundary. 

Note, that the mean UP boundary position for the type~II $\alpha = 1$ simulation shown in Fig.~\ref{radial_profile} and in Table~\ref{fractal_dim} is at $r/r_\mathrm{spot} = 0.64$ where the visual position of the boundary in Fig.~\ref{spots_int} is very close to the actual sunspot boundary. This is caused by a number of bright structures within the umbra that results into $I_\mathrm{c} > 0.5$ when averaged along iso-contours of relative radial positions. This is also the reason why we obtain significantly larger $B_\mathrm{ver}$ on the mean position of the UP boundary than on the actual UP boundary.

\section{Plasma \texorpdfstring{$\beta$}{beta} and equipartition field strength}

Using the results of the simulation, we can compute plasma $\beta$ and equipartition magnetic field strength $B_\mathrm{eq}$ using
\begin{equation}
    \beta = \frac{8 \pi P_\mathrm{gas}}{B^2}, 
\end{equation}
\begin{equation}   
    B_\mathrm{eq} = (4 \pi \rho)^\frac{1}{2} v,
\end{equation}
where $P_\mathrm{gas}$ is the total gas pressure, $B$ is the total magnetic field strength, $\rho$ is the density, and $v$ is the velocity. All these physical parameters are direct output of the simulation run. 

 The $B_\mathrm{eq}$ cannot be determined from observations as we can access only the line-of-sight component of $v$. In principle, we can evaluate $\beta$ using the $B$ value determined by the inversion code. However, it would be highly questionable to assign a unique value of $P_\mathrm{gas}$ to it since the total gas pressure is decreasing rapidly in the range of optical depths where the inverted lines are most sensitive to $B$ values \citep[$-2 < \log(\tau) < -1$,][]{Cabrera:2005}.

In Fig.~\ref{forces_tau1}, we show the resulting values of plasma $\beta$ and the ratio of $B_\mathrm{eq}/B$ at the continuum formation height for different simulations. In the case of the simulations with potential field extrapolation (type~I and type~II, $\alpha = 1$), we can clearly distinguish different regimes of plasma $\beta$ in different areas of the simulation box. In sunspot umbrae, the magnetic pressure dominates over the gas pressure. In sunspot penumbrae, the values of $\beta$ are around 1. When studied in detail, we find plasma $\beta > 1$ in bright penumbral filaments and $\beta < 1$ in the regions between them. In quiet Sun regions, the gas pressure dominates over the magnetic pressure and $\beta$ is around 1 only in magnetised regions that typically appear as small dark areas in continuum intensity maps (see Fig.~\ref{spots_int}). In the case of the non-potential simulation (right map in Fig.~\ref{forces_tau1}), the significant difference in the value of $\beta$ is the inner penumbra, where the magnetic pressure dominates over the gas pressure even in bright penumbral filaments.

\begin{figure}[!t]
 \sidecaption
 \includegraphics[width=\linewidth]{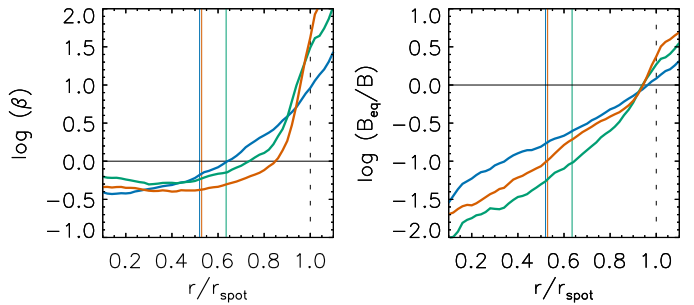}
 \caption{Radial profiles of $\log(\beta)$ (left) and $\log(B_\mathrm{eq}/B)$ (right) at the continuum formation height. Colour coding and the vertical lines are analogous to Fig.~\ref{radial_profile}. The horizontal lines mark the threshold where $\beta = 1$ (left) and $B_\mathrm{eq}/B = 1$ (right).}
 \label{radial_profile_forces}
\end{figure}

In the lower row of Fig.~\ref{forces_tau1}, we show the ratio of $B_\mathrm{eq}/B$. In the case of this parameter, $B$ dominates over $B_\mathrm{eq}$  both in the sunspot umbrae and penumbrae and this behaviour is independent on the type of the simulation. $B \sim B_\mathrm{eq}$ at the outer penumbral boundary with $B_\mathrm{eq}$ thus defining the sunspot border.  This was already suggested in observational studies describing the magnetic field properties at the sunspot boundaries \citep{Wiehr:1996, Kalman:2002}. $B_\mathrm{eq}$ dominates over $B$ in the quiet Sun regions, again with the exception of the small and concentrated magnetic patches.

In Fig.~\ref{radial_profile_forces}, we show the radial profiles of $\beta$ and $B_\mathrm{eq}/B$ in the sunspots displayed in Fig.~\ref{forces_tau1}. These radial profiles also show that, on average, the gas pressure starts to dominate over the magnetic pressure in the inner penumbra for type~I simulation (blue line) and type~II $\alpha = 1$ simulation (green line) and only in the outermost penumbra in the case of type~II $\alpha = 2.5$ simulation (red line). The mean position of the boundary between the penumbra and the quiet Sun matches well with the location where $B_\mathrm{eq}$ gets stronger than $B$ for all of the simulated spots, i.e., all the radial profiles of  $\log(B_\mathrm{eq}/B)$ cross zero at $r/r_\mathrm{spot} \sim 0.95$.   

\begin{figure}[!t]
 \sidecaption
 \includegraphics[width=\linewidth]{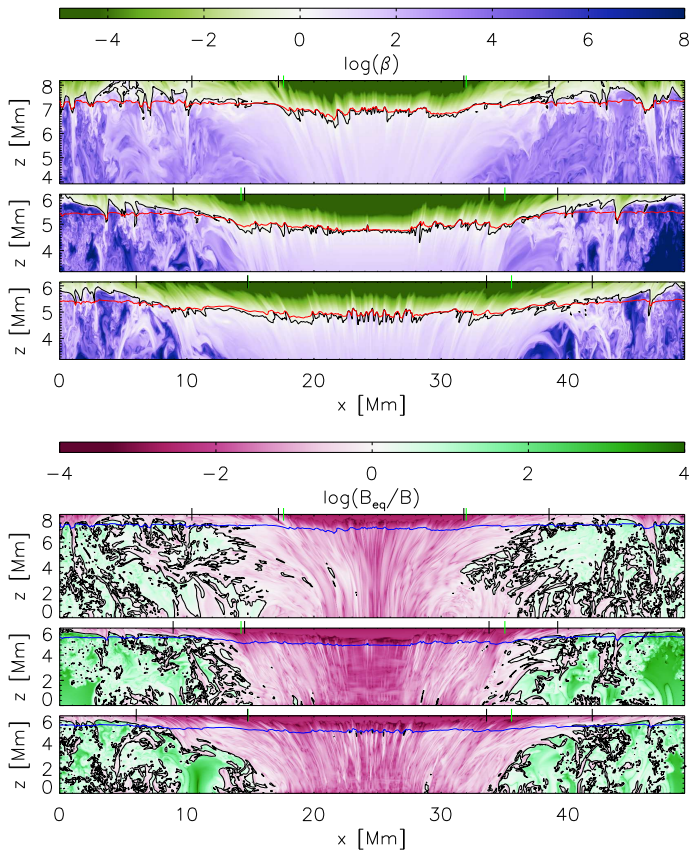}
 \caption{Stratification of $\log(\beta)$ (top panels) and $\log(B_\mathrm{eq}/B)$ (bottom panels) along the cuts displayed in Fig.~\ref{forces_tau1}. The red (top) and blue (bottom) lines mark the $\tau =1$ surface. The black lines mark the zero value of $\log(\beta)$ and $\log(B_\mathrm{eq}/B)$. The short vertical lines crossing the upper edges of the panels mark the positions, where the cuts cross the contours displayed in Fig.~\ref{forces_tau1}. Note that only the upper part of the simulation domain is displayed in the case of $\log(\beta)$.}
 \label{forces_cuts}
\end{figure}

In Fig.~\ref{forces_cuts}, we show the vertical stratification of $\beta$ and $B_\mathrm{eq}/B$ along cuts through the centre of the simulated sunspots. It is clear that the gas pressure dominates over the magnetic pressure in most of the sub-photospheric layers of the simulation domains. Only in regions of the umbrae, there is a thin layer below the continuum formation height where the magnetic pressure is larger than the gas pressure, i.e., the black lines in the upper panels of Fig.~\ref{forces_cuts} are located just below the red lines. 

The vertical cuts showing $B_\mathrm{eq}/B$ nicely illustrates that the  magnetopause between the sunspot and the surrounding plasma coincides with the transition from super- to sub-equipartition field strength. Since $B_\mathrm{eq}/B$ is less dependent on height within the sunspot than $B$, this quantity is better suited than $B$ (Fig.~\ref{mag_cut}) to outline the shape of the spot flux rope. Locations where this magnetopause cross the $\tau =1$ surface correspond to the outer boundaries of the simulated sunspots and confirm the conclusions based on Figs.~\ref{forces_tau1} and~\ref{radial_profile_forces}.

\section{Discussion and conclusions}
\label{discussion}

We compared the magnetic properties of a set of simulated sunspots with an observed sunspot of comparable magnetic flux. Despite the visual similarity of type~II $\alpha \geq 1.5$ simulations with observed sunspots, it turns out that the magnetic properties in these simulations do not correspond to those in observed sunspots. Most notably, the magnetic field is too horizontal on the UP boundaries of simulated sunspots ($\gamma \sim 60^\circ$, Table~\ref{UP_properties}) where we find maximal values of $\gamma$ around $45^\circ$ in observed sunspots \citep{Jurcak:2018}. While these simulation setups do produce the most extended penumbrae, this comparison indicates that creating a penumbra through imposing a strong horizontal field from above is not a viable approach.

Simulations with potential field extrapolation (type~I and type~II $\alpha =1$) are closer to the observed sunspots in terms of the magnetic properties on their UP boundaries. In the case of the type~II $\alpha =1$ simulation, we found the best match of $B$, $\gamma$, and $B_\mathrm{ver}$ with the observed sunspot. However, this particular simulation does not have any distinct penumbra. 

The extent of the penumbra for the simulations with potential field extrapolation is clearly the consequence of the subsurface structure of the magnetic flux tube in the simulation domain. The type~I simulation has a very concentrated footpoint at the bottom boundary and thus the field expands significantly with height in the simulation domain, shaping a sufficiently inclined magnetopause that allows for an extended penumbra (see Fig.~\ref{mag_cut} and bottom panels in Fig.~\ref{forces_tau1}). However, the magnetic field strength of the type~I simulation is too high in the photosphere for a sunspot of such magnetic flux (see Fig.~\ref{radial_profile}). Note that the type~I simulation has another discrepancy with the observed sunspots, the Evershed flow is observed only in a minority of the penumbral filaments and in a majority of the penumbral filaments we observe a counter-Evershed flow. This is also the case for type~II $\alpha =1$ simulation that does not show long penumbral filaments carrying Evershed flows.

Another aspect of observed sunspots, that is not matched by the simulations, is the relative position of the UP boundary with respect to the sunspot radius (see Table~\ref{fractal_dim}). In the case of the observed sunspots $r/r_\mathrm{spot} \sim 0.4$. In the case of the simulations with potential field extrapolation, we obtain UP boundaries at $r/r_\mathrm{spot}$ of 0.52 and 0.64 for type~I and type~II $\alpha = 1$, respectively. Therefore, tuning the initial setup of the simulation to obtain the mean position of UP boundary at $r/r_\mathrm{spot} = 0.4$ can restrict the subsurface structure of the magnetic flux tube. 

Another discrepancy between the simulated and observed sunspots is the complexity of the UP boundary. In Table~\ref{fractal_dim}, we show that the observed sunspot has significantly less corrugated shape of the UP boundary than any of the simulated sunspots. This property of the simulated sunspots partially accounts for the discrepancies between simulated and observed sunspots. When we disregard the fine shape of the UP boundaries and compare the values of $B_\mathrm{ver}$ at the mean position of the UP boundary, we find that the observed value of 1.82~kG is comparable to the values found in type~I and type~II $\alpha =1$ simulations of 1.71~kG and 1.78~kG, respectively. 

The run of a simulation that covered 100 hours of solar time and a box depth of 18~Mm did not have any clear impact on the corrugation of the UP boundary \citep{Rempel:2015}. Also, simulations with enhanced spatial resolution did not produce a smooth UP boundary \citep{Rempel:2012}. However, the simulations with high spatial resolution were not run for a long time nor with enhanced depth of the simulation domain, so we cannot yet exclude that such simulations would produce UP boundaries comparable in terms of their smoothness to the observed ones (after degradation of the simulation to the spatial resolution of the observations).

We used the simulations also to assess the reliability of our simple inversion scheme. In Figs.~\ref{overview_maps} and~\ref{scatter_plots}, we compared the results of the inversion of the Stokes profiles synthesised from the type~I simulation with the actual values in the simulation domain. Especially in the regions where the UP boundary is located ($B > 2$~kG, $20^\circ < \gamma < 70^\circ$), the inversions give reliable results. Assuming that the simulated atmosphere is of comparable complexity to the actual solar photosphere, we can conclude that the mean values of $B_\mathrm{ver}$ derived from analyses of Hinode spectropolarimetric data are reliable. 

We investigated the role of plasma $\beta$ and $B_\mathrm{eq}$ on the convective motions in the simulations as these parameters cannot be determined reliably from observations. In the case of plasma $\beta$, we found that, at the solar surface, sunspot umbrae are dominated by magnetic pressure, quiet Sun regions by gas pressure, and in sunspot penumbrae $\beta \sim 1$ (Fig.~\ref{forces_tau1}). On average, the gas pressure starts to dominate over the magnetic pressure in the inner penumbra in case of the simulations with potential field extrapolation (Fig.~\ref{radial_profile_forces}). 
$\beta = 1$ appears to outline the UP boundary, yet, we assume that this is just a consequence of the presence of convective structures (penumbral filaments) and not the cause for their appearance. An analysis on the basis of the Gough-Tayler criterion (cf. Eq. (1)) to understand the cause of convective instability in the penumbra is being developed.

The analysis of the equipartition field strength and its ratio to the magnetic field strength shows that the magnetopause  coincides  with  the  transition from  super-  to  sub-equipartition  field  strength (Fig.~\ref{forces_cuts}). Where this magnetopause crosses the solar surface, we observe the boundary of the sunspot (Figs.~\ref{forces_tau1} and~\ref{radial_profile_forces}).  In other words, sunspots outer boundary is defined by magnetic field strengths of value $B_\mathrm{eq}$.

\begin{acknowledgements}
JJ acknowledges the support from the Czech Science Foundation grant 18-06319S. This research has made use of NASA's \href{https://ui.adsabs.harvard.edu/}{Astrophysics Data System}. MS thanks HAO/NCAR's visitor program for financing his stay at HAO. This material is based upon work supported by the National Center for Atmospheric Research, which is a major facility sponsored by the National Science Foundation under Cooperative Agreement No. 1852977. We would like to acknowledge high-performance computing support from Cheyenne (doi:10.5065/D6RX99HX) provided by NCAR's Computational and Information Systems Laboratory, sponsored by the National Science Foundation. Some of the figures within this paper were produced using IDL colour-blind-friendly colour tables \citep[see][]{pjwright}.

\end{acknowledgements}

\bibliographystyle{aa}
\bibliography{manuscript}

\end{document}